\documentclass[useAMS,usenatbib]{mn2e}
\usepackage{latexsym}
\usepackage{graphicx}
\usepackage{amssymb}
\usepackage{longtable}
\usepackage{multirow}
\usepackage{epsf}
\usepackage{pbox}
\usepackage{times}
\DeclareGraphicsExtensions{.ps,.pdf,.png,.jpg}

\title[HD 189733\lowercase{b} atmospheric structure retrievals from transit spectroscopy]{Optimal Estimation Retrievals of the Atmospheric Structure and \\ Composition of HD 189733\lowercase{b} from Secondary Eclipse Spectroscopy}
\author[J. -M. Lee, L. N. Fletcher and P. G. J. Irwin]{J. -M.~Lee$^1$\thanks{E-mail: lee@atm.ox.ac.uk}, L. N.~Fletcher$^1$ and P. G. J.~Irwin$^1$ \\
$^1$Atmospheric, Oceanic, and Planetary Physics, University of Oxford, Oxford OX1 3PU}
\date{Accepted; Received; in original form}

\pagerange{\pageref{firstpage}--\pageref{lastpage}} \pubyear{2011}

\def\LaTeX{L\kern-.36em\raise.3ex\hbox{a}\kern-.15em
    T\kern-.1667em\lower.7ex\hbox{E}\kern-.125emX}

\begin{document}

\label{firstpage}

\maketitle

\begin{abstract}
\noindent Recent spectroscopic observations of transiting hot Jupiters have permitted the derivation of the thermal structure and molecular abundances of H$_{2}$O, CO$_{2}$, CO, and CH$_{4}$ in these extreme atmospheres. Here, for the first time, we apply the technique of optimal estimation to determine the thermal structure and composition of an exoplanet by solving the inverse problem. The development of a suite of radiative transfer and retrieval tools for exoplanet atmospheres is described, building upon a retrieval algorithm which is extensively used in the study of our own solar system. First, we discuss the plausibility of detection of different molecules in the dayside atmosphere of HD 189733b and the best-fitting spectrum retrieved from all publicly available sets of secondary eclipse observations between 1.45 and 24 $\mu$m. Additionally, we use contribution functions to assess the vertical sensitivity of the emission spectrum to temperatures and molecular composition. Over the altitudes probed by the contribution functions, the retrieved thermal structure shows an isothermal upper atmosphere overlying a deeper adiabatic layer (temperature decreasing with altitude), which is consistent with previously-reported dynamical and observational results. The formal uncertainties on retrieved parameters are estimated conservatively using an analysis of the cross-correlation functions and the degeneracy between different atmospheric properties. The formal solution of the inverse problem suggests that the uncertainties on retrieved parameters are larger than suggested in previous studies, and that the presence of CO and CH$_{4}$ is only marginally supported by the available data. Nevertheless, by including as broad a wavelength range as possible in the retrieval, we demonstrate that available spectra of HD 189733b can constrain a family of potential solutions for the atmospheric structure.
\end{abstract}


\section{Introduction}

The vertical structure of an exoplanetary atmosphere can be derived from transmission and emission spectroscopy when a planet transits its host star. The atmospheric properties of HD 189733b have been extensively investigated using transit spectroscopy observations from both space and ground-based telescopes. Transmission spectra acquired during primary transits, where light from the host star is filtered through the upper layers of an exoplanetary atmosphere, have been measured in the near-, mid- and far-infrared (NIR, MIR and FIR, respectively) by both the $Hubble$ $Space$ $Telescope$ ($HST$) \citep{swa08,sin09} and $Spitzer$ telescope \citep{tin07a,bea08,des09}.  $Spitzer$ mid- and far-infrared (MIR and FIR) broadband photometry was used to deduce a high abundance of H$_{2}$O in the terminator region of this planet \citep{tin07a,swa08,bea08,des09}.  In addition, a strong absorption due to CH$_{4}$ was reported in the near-infrared (NIR) using the NICMOS (Near Infrared Camera and Multi-Object Spectrometer) spectrophotometry on the $HST$ \citep{swa08}. However, the accuracy of transit spectroscopy and photometry has been called into question \citep[see the review by][]{sea10}, and most recently \citet{gib11} claimed that instrumental systematics limit our ability to constrain atmospheric CH$_{4}$ from these $HST$/NICMOS data as well as CO$_{2}$ and CO. A spectral feature at 4.5 $\mu$m can be explained by either of the carbon bearing molecules CO$_{2}$ or CO, although it is still unclear which due to the degeneracy of the solution \citep{des09,for10}. At visible wavelengths, strong lines of the alkali metal sodium were reported using observations with the high resolution ground spectrograph of the $\it{Hobby}$--$\it{Eberly}$ Telescope \citep{red08}, while featureless spectra detected by the $HST$/ACS \citep{pon08} and $HST$/STIS \citep{sin11} are thought to be caused by thick atmospheric hazes. 

Observations of secondary transit spectra of HD 189733b by $Spitzer$ broadband photometry, when emission spectra from the planet's dayside is detected before and after the planet is eclipsed, have suggested that H$_{2}$O is also abundantly present in the dayside atmosphere \citep{cha08,dem06}, forming the main features of the emission spectrum. The H$_{2}$O-rich atmosphere was confirmed via $Spitzer$ IRS (Infrared Spectrograph) dayside spectroscopy \citep{gri08}. $HST$/NICMOS dayside observations indicated a high abundance of CO$_{2}$, CO, and CH$_{4}$ \citep[][hereafter S09]{swa09}, supporting the conclusions of their primary transit spectrophotometry. With all measurements, each of these studies suggested that an atmospheric thermal profile lacking a thermal inversion (i.e. without a warm stratosphere) could best explain their observations.  Finally, a decreasing temperature with altitude between 0.01-1.0 bar suggested the presence of a troposphere \citep[S09;][hereafter MS09]{mad09}. This work revisits the secondary transit spectroscopy of this hot Jupiter to assess the family of atmospheric structures consistent with these measurements.

Various techniques have been used to investigate the atmospheric constituents and structure from transit spectroscopy. Infrared dayside spectra during secondary eclipses measure the thermal emission of the dayside of the exoplanet, which depends mostly on the vertical temperature structure, while primary eclipse observations are more sensitive to the absolute temperature at the terminator due to its scale height dependence \citep{bro01,tin07b}. The retrieval of atmospheric properties from remote sensing measurements has been developed over many decades \citep{goo89} and has become a common tool for the study of planetary atmospheres. Measured radiances (either photometry or spectroscopy) can be compared to synthetic spectra from radiative transfer modelling so as to determine the atmospheric structure and composition. Based on the Bayesian approach, an optimal estimation retrieval model solves an ill-constrained inversion problem using an iterative scheme to maximise the probability of solutions to the available data (the maximum $a$ $posteriori$ solution) \citep{rod00}. Previous studies of the terminator and dayside spectrum, such as those by \citet{tin07a}, \citet{swa08} and S09, have utilised forward modelling to determine a best-fitted spectrum to the data, and have constrained the range of molecular abundances of transiting exoplanets by synthesising spectra based on theoretical pressure-temperature ($P$-$T$) profiles. An alternative forward modelling approach used by a number of studies \citep[MS09;][]{sin08,mad10,ste10,mad11,mad112} used freely roaming $P$-$T$ profiles and molecular abundances in a parameterised space, where calculated spectra were again compared with observations in terms of the goodness-of-fit, and their numerous runs were able to constrain $P$-$T$ profiles and atmospheric compositions. Although these forward modelling techniques provide valuable insights, their methods are based on line-by-line radiative transfer models, which are slow when large numbers of synthetic spectra are to be calculated over a wide wavelength range. Moreover, the huge degeneracies between the different model parameters were not explored in detail.

The NEMESIS optimal estimation retrieval algorithm \citep{irw08} uses the correlated-$k$ technique \citep{lac91} in its radiative transfer model, which rapidly integrates synthetic spectra using $k$-distribution tables pre-calculated from line databases \citep{goo89}. The combination of the correlated-$k$ forward model, which is orders of magnitude faster than a standard line-by-line model, with an optimal estimation retrieval scheme \citep{rod00} has been used to successfully investigate planetary atmospheres in our own solar system. In this study we apply this rapid retrieval architecture to exoplanets for the first time, which allows us to formally address the uncertainties and degeneracies inherent in previous studies. Moreover, we use the covariance matrices calculated via the inversion method to quantify the correlations between the derived $P$-$T$ profile and atmospheric composition. Consequently, this study will show how the characteristics of these extreme exoplanet atmospheres can be deduced from spectroscopic measurements and highlight the limitations of the datasets available today.

In this report, the modelling method used to retrieve the atmospheric properties from the dayside spectroscopy and photometry of HD 189733b is described in Section 2. All available measurements used to constrain atmospheric properties are described in Section 3 and the best-fitting solution to these measurements is discussed in Section 4. Section 5 demonstrates the validity of the retrieval scheme, and quantifies the degeneracy of the solution. In Section 6 we discuss the implications of our results and arrive at our conclusions.

\section{Modelling}

The atmospheric thermal structure and composition are derived using an optimal estimation retrieval algorithm, NEMESIS \citep{irw08}. Instead of simply minimising the deviation between data and model by comparing the observations to millions of individually calculated synthetic spectra (the forward model), this algorithm instead solves the inverse problem using an iterative scheme.  The retrieval problem is ill-posed (i.e. the number of unknowns is greater than the number of measurements) and ill-conditioned (i.e. small changes in the measurements can potentially have a disproportionately large effect on the solution) so there is no single unique solution.  Instead, physically-realistic solutions and a range of uncertainty must be defined for the family of possibilities.  Following \citet{rod00}, we derive the optimal atmospheric state ($\widehat{\mathbf{x}}$) via the inclusion of additional \textit{a priori} constraints, in the form of an $a$ $priori$ state vector, $\mathbf{x_a}$, with error covariance matrix, $\mathbf{S_a}$.  Thus successive iterations are used to minimise the cost function, $\phi$ \citep{irw08,rod00}:
\[
\phi=(\mathbf{y}-\mathbf{F(\widehat{x})})^{T}\mathbf{S_\epsilon}^{-1}(\mathbf{y}-\mathbf{F(\widehat{x})})+(\mathbf{\hat{x}}-\mathbf{x_a})^T\mathbf{S_a^{-1}}(\mathbf{\widehat{x}}-\mathbf{x_a})
\]
where $\mathbf{F(x)}$ is the forward model (i.e. the spectrum calculated for the trial atmosphere); $\mathbf{x}$ is the state vector (which may be a continuous atmospheric profile of temperature or composition, or a scaling factor of a known profile) and $\mathbf{y}$ is the measurement vector with measurement error covariance matrix, $\mathbf{S_{\epsilon}}$.  The first term requires that the synthetic spectra must give a close fit to the measured spectrum, the second term requires that the solution remains physically realistic.   \citet{rod00} shows that the optimal estimation solution for the state vector is given by:
\[
\widehat{\mathbf{x}} = \mathbf{x_a}+\left(\mathbf{K^TS_\epsilon^{-1}K+S_a^{-1}} \right)^{-1}\mathbf{K^{T}S_\epsilon^{-1}}(\mathbf{y-Kx_a}) \\
\]
where $\mathbf{K}$ is the matrix of functional derivatives (i.e. the rate of change of radiance with each model parameter). The optimal state in the algorithm is accomplished after a moderate number of iterations (approximately 10--20), using a scheme based on the Marquardt-Levenberg principle \citep{lev44,mar63}. The algorithm calculates synthetic spectra using the correlated-$k$ approximation method, which has been shown to be fast, reliable, and sufficiently accurate compared to line-by-line calculations. We compared spectra computed by line-by-line and correlated-$k$ models, and found that the mean difference between the two techniques is less than 5 per cent, which is considerably smaller than the range of uncertainties on available exoplanet measurements.

In the $k$-distribution method, the absorption spectrum over an interval is sorted in order of increasing absorption and the fraction of the interval with absorption less than a certain value $k(g)$ is represented in terms of the fraction of the interval $g$. Since $k(g)$ is a smoothly varying function of $g$, it may be integrated with relatively few quadrature points $N$ to determine the mean transmission over the interval as: 

\[
\bar{T}(m)\simeq\sum\limits_{i=1}^N e^{-k_{i}m}\Delta g_{i},
\]

\noindent where $m$ is the amount of a molecule; and $k_{i}$ and $\Delta g_{i}$ are the $k$-coefficients and quadrature weights at each quadrature point \citep{irw08}. The $k$-coefficients are calculated from line data for a range of temperatures and pressures expected in the atmosphere in advance of the retrieval to enable rapid calculation of the transmission during each iteration. The molecular line lists used in this study were taken from HITEMP2010 \citep{rot10} for H$_{2}$O, the Carbon Dioxide Spectroscopic Databank (CDSD-1000) \citep{tas03} (also used for HITEMP2010) for CO$_{2}$, HITEMP1995 \citep{rot95} for CO, and the Spherical Top Data System (STDS) \citep{wen98} for CH$_{4}$. The absorption coefficients for the extreme atmospheric temperatures found within exoplanetary atmospheres are continuously being improved, so the conclusions of this study are considerably limited by the quality of the available spectroscopic data. We have made use of the best available spectroscopic parameters at the time of writing, and our line database will be updated as new sources of data become available. 

The combination of the optimal estimation retrieval scheme and the correlated-$k$ method enhances the efficiency of the retrieval process in terms of time and computational resources. Furthermore, NEMESIS calculates the matrix of the partial derivatives of radiances at each wavelength with respect to each retrieved variable, which are called the $functional$ $derivatives$ (or $Jacobians$), in order that the contribution of different atmospheric parameters at each wavelength can be easily interpreted by comparing the elements of this matrix. 

Our $a$ $priori$ dayside atmosphere of HD 189733b extends from 10$^{-9}$ to 10 bar to capture the full atmospheric range of potential spectral contributions. For initial modelling we assumed that all species were well-mixed throughout the atmosphere and the molecular abundance was defined in terms of a single scaling parameter. This is because the retrieval of a continuous profile of composition would be under-constrained considering the small number of data points and the low spectral resolutions available, leading to non-physical oscillations in the retrieved profile. The $a$ $priori$ estimate for the abundance scaling parameter is assumed to have a large uncertainty so that retrieved values are not weighted by initial guesses since a simple scaling parameter already includes vertical smoothing. The $a$ $priori$ for temperature, however, is assumed to have a continuous profile and the assumed error on the $a$ $priori$ profile was adjusted to achieve the optimal balance between the quality of the fit to the measured data and the vertical smoothing. This will be further discussed in Section 5.3.1. 

As an important source of absorption in exoplanetary atmospheres, collisional-induced absorption (CIA) between principle gases are included in the model atmosphere. We consider the interactions between H$_{2}$--H$_{2}$ and H$_{2}$--He and the coefficients are taken from \citet{bor89}, \citet{bor892}, \citet{bor90}, \citet{zhe95}, \citet{bor97}, and \citet{bor02}. The mole fractions of H$_{2}$ and He are assumed to be related to the fractions of atomic H and He, which are close to the typical solar value of 0.91 and 0.0887 each \citep{bur99}.  The implications of this assumption will be tested in Section 5.5.

Scattering by clouds and hazes has been reported from $HST$ observations at visible wavelengths, yielding a featureless transmission spectrum of HD 189733b \citep{pon08,sin11}. Recently, \citet{hen11} showed that the effect of scattering by clouds and hazes modifies the inferred temperature profile, and that an isothermal temperature structure above an adiabatic troposphere could also be caused by a cloud-top or haze layer. We consider, however, that the inclusion of scattering clouds and hazes in our retrieval model is beyond the scope of the present study for a number of reasons: (i) the scattering properties of possible clouds and hazes is insufficiently known; (ii) adding such particles would greatly enlarge an already large and poorly constrained parameter space and; (iii) modelling scattering processes will significantly increase the computation time. In summary, the addition of scattering would increase the complexity of the model and does not appear to be warranted by the dayside emission spectra studied here.  We intend to assess the likely effects of scattering on our modelling of $hot Jupiters$ in a follow-up study.

\begin{figure}
\hspace{-0.3cm}
\includegraphics[width=9cm]{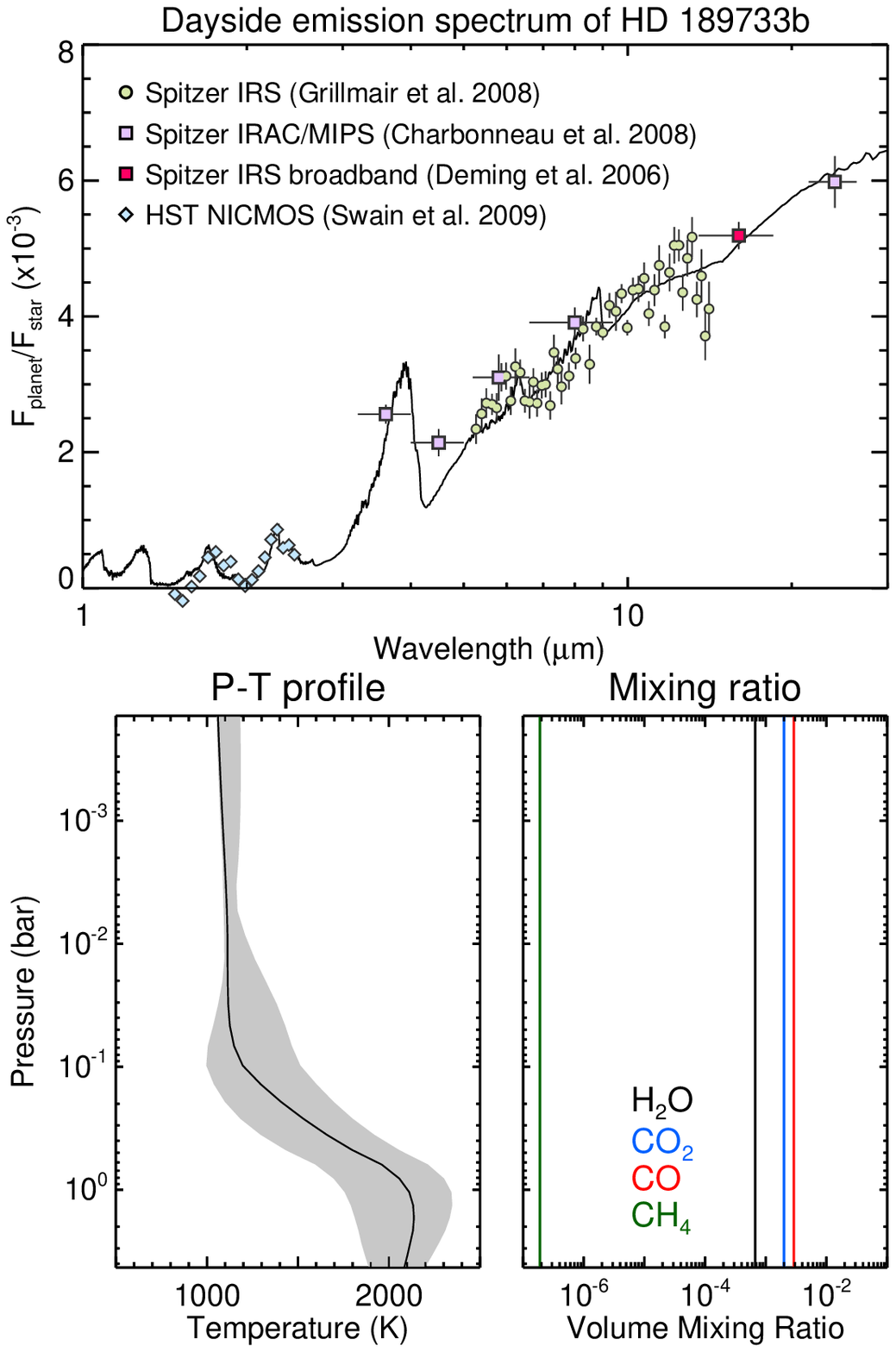}
\caption{The best-fitting spectrum and atmospheric properties retrieved by the NEMESIS algorithm. In the top panel, the best-fitting dayside emission spectrum of HD 189733b is displayed as a black line. The green, purple with red, and green symbols are the measured planet-star flux ratio from the $Spitzer$ IRS spectroscopy \citep{gri08}, the $Spitzer$ broadband photometry \citep{cha08,dem06}, and the $HST$/NICMOS spectrophotometry \citep{swa09} (cf. Section 4). The retrieved best-fitting $P$-$T$ profile in the bottom-left panel is shown in a black solid line with grey-coloured uncertainties due to the molecular degeneracy (cf. Section 5.3). In the bottom-right panel, retrieved best-fitting molecular mixing ratios for H$_{2}$O, CO$_{2}$, CO, and CH$_{4}$ are demonstrated with different colours. Each abundance is assumed to be fixed with height in order to minimise the number of retrieval variables (cf. Section 5.4). }
\label{f1}
\end{figure}

\section{Data}
To retrieve the $P$-$T$ profile and compositional abundances, we used three measurement sets of the secondary eclipse for HD 189733b, representing all of the available infrared dayside eclipse measurements at the present time, ranging over a wide wavelength range from 1.45 $\mu$m to 24 $\mu$m: (i) eighteen $HST$/NICMOS \citep{swa09} channels covering the range 1.45--2.5 $\mu$m; (ii) forty seven $Spitzer$ IRS \citep{gri08} channels covering the range 5--14.5 $\mu$m and one IRS photometry \citep{dem06} channel at 16 $\mu$m; and (iii) five $Spitzer$ IRAC (Infrared Array Camera) and MIPS (Multiband Imaging Photometer for $Spitzer$) \citep{cha08} photometry channels at 3.6, 4.5, 5.8, 8.0, and 24 $\mu$m. The measurement errors on the three observational datasets are directly obtained from the studies referenced above. To integrate the binned flux in each channel, we use (i) filter widths taken from the literature for $Spitzer$ broadband channels \citep{faz04,rie04,dem06}; (ii) $HST$/NICMOS widths of $\sim$10 nm at 2 $\mu$m; and (iii) $Spitzer$/IRS widths of $\sim$100 nm at 8 $\mu$m. The reference stellar spectrum for HD 189733 is taken from the Kurucz grid model\footnote{http://kurucz.harvard.edu/stars/HD189733/}. 

Despite the substantial efforts towards finding and reducing the errors from the data, uncertainties on the given datasets still remain and are widely distributed over the wavelengths due to various error sources. First of all, techniques to decorrelate a transit light curve from the combined light of a planetary system are not consistent each other [e.g. \citet{swa08} vs. \citet{gib11}]. We will discuss the effects of underestimated errors on the dayside spectra in Section 4.1.  Also, systematic errors inherent in the observations themselves could be a strong error source. These two facts may cause the planet-star flux ratio to be dependent on the data reduction process. Non concurrent observations for different wavelengths could cause large variations in the measured flux between observations if there is any significant temporal variability in the atmosphere of HD 189733b. Therefore, all these sources of potential error could lead to substantial inconsistencies between the datasets taken from different studies, instruments, and observation times. Although we do not intend to revisit the decorrelation procedures with this analysis, we aim to show the limitations of the spectra available at the present epoch under the assumption that the measurements are an accurate representation of the mean planetary flux.

\section{Best-fitting dayside spectrum of HD 189733\lowercase{b}}
\begin{figure*}
\hspace{0.4cm}
\includegraphics[width=16cm]{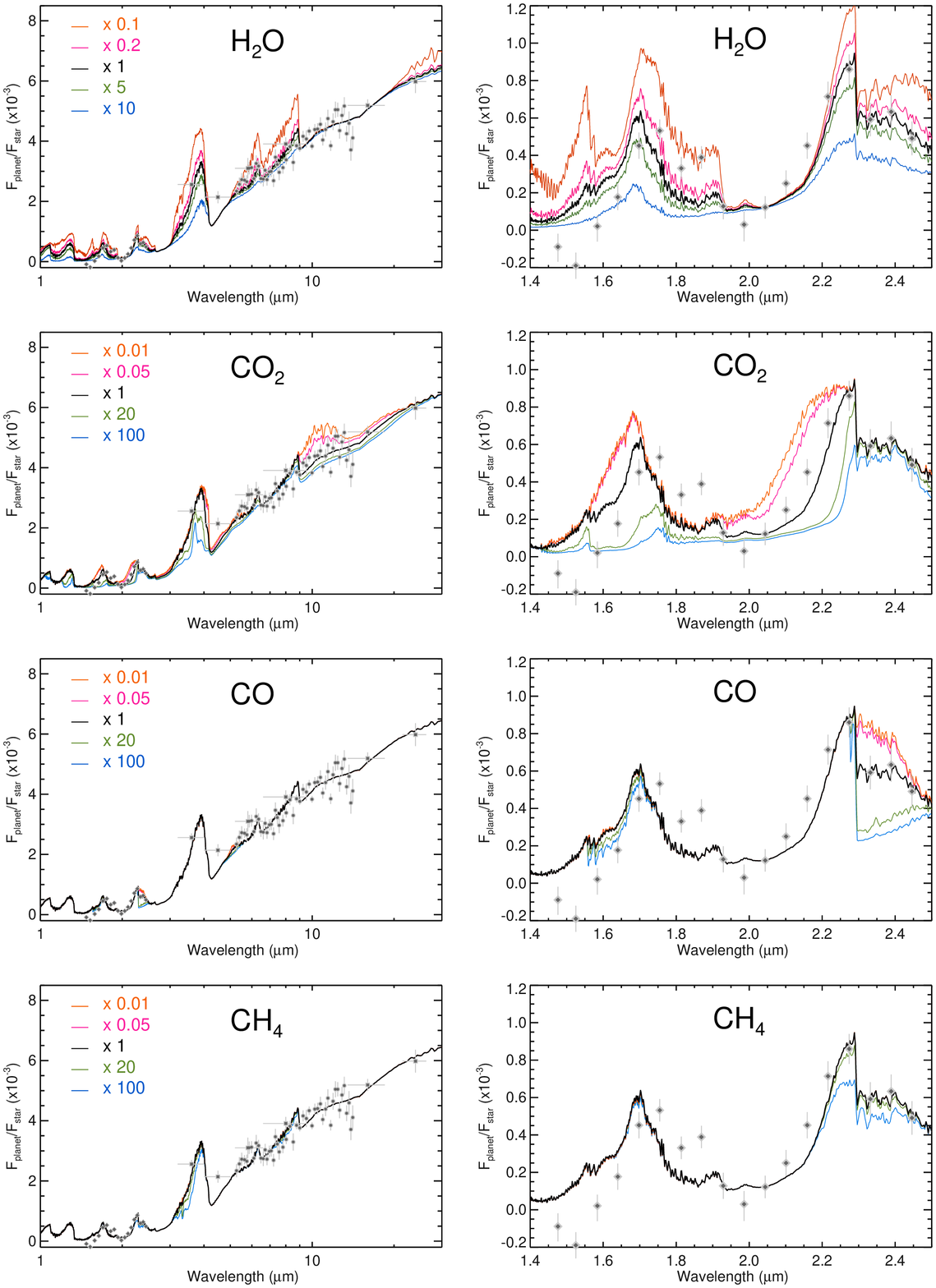}
\caption{Fitted dayside emission spectra of HD 189733b. All available measurements are shown in grey symbols. The best-fitting spectrum is displayed as a black line in all figures. We also show calculated spectra with various molecular abundances to understand the contributions of different molecules to the best-fitting spectrum. For H$_{2}$O, molecular abundances are varied by 0.1, 0.2, 5.0 and 10.0 times the abundance leading to the best-fitting spectrum, and by 0.01, 0.05, 20.0 and 100.0 times for the other molecules.}
\label{f2}
\end{figure*}

\subsection{Significance test ($F$--test)}
The molecular composition of the dayside atmosphere of HD 189733b is generally accepted to be a mixture of H$_{2}$O, CO$_{2}$, CO, and CH$_{4}$. These molecules have been discovered based on various spectroscopic measurements, listed in Section 3. However, given the large uncertainties on the available data and the degeneracy between retrieved parameters, the existence of some of these species is called into question.  We therefore set out to statistically test whether the presence of each molecule is truly warranted by the data.  Using a statistical significance test known as an $F$--test \citep{sne89}, we evaluated the change in confidence level between a range of simple and complex models.  We successively added more complexity to the model (i.e. added molecules to the H$_{2}$/He atmosphere to increase the number of degrees of freedom) and evaluated the improvement in the goodness-of-fit parameter ($\chi^2$) for the best-fitting spectrum.  The $F$--test was then used to assess whether the inclusion of a particular molecule was required to fit the data.

The $F$--test indicates that an atmospheric composition for HD 189733b containing H$_{2}$O and CO$_{2}$ are highly plausible at the $>$99.98\% confidence level.  The presence of these two molecules is required to obtain a reasonable fit to the available data.  In addition,  we find insignificant confidence levels (i.e. $<$95\%) when we increase the complexity of the model by adding CO, CH$_4$ or a combination of the two to the simple model containing H$_{2}$, He, H$_{2}$O and CO$_{2}$.  In other words, CO and CH$_{4}$ provide negligible enhancements to the fitting quality and are not required to fit the dayside emission measurements to within the stated error bars (although upper limits on these species can certainly be derived).

Additionally, we consider the implications of \citet{gib11}, who claimed larger uncertainties on the $HST$/NICMOS data by re-processing its transmission spectrum \citep{swa08}. If the same conclusions are applicable to the secondary eclipse emission spectrum, then we must similarly increase the measurement error on the NICMOS data by a factor of five.  This leads to even smaller confidence levels on the more complex models, and the solutions are found to be even more degenerate.  In this case, evaluating the $F$--test significance using a variety of models still suggests that both H$_{2}$O and CO$_{2}$ are required to reproduce the dayside emission spectra ($>$99.29\% confidence), but makes the presence of CO and CH$_4$ even more uncertain.  

We therefore conclude that, irrespective of the uncertainties on the $HST$/NICMOS data, current datasets are unable to provide detections of CO and CH$_4$ on the dayside of HD 189733b with any sort of statistical certainty.  Nevertheless, upper limits on the abundances of these molecules can be obtained (see Section 5), and all four molecules will be included in our subsequent study.

\subsection{Best-fitting solution}
Using the NEMESIS algorithm, we retrieve the best-fitting dayside spectrum of HD 189733b, incorporating both the $Spitzer$ and $HST$ observations as stated in Section 3. Fig.~\ref{f1} demonstrates the best-fitting spectrum to these measurements, and the retrieved atmospheric $P$-$T$ profile and molecular abundances for H$_{2}$O, CO$_{2}$, CO, and CH$_{4}$. Each panel will be described below.

Fig.~\ref{f2} shows the contributions from the four main gases included in our model. This figure also shows the wavelength ranges where the molecular contributions are distributed by co-plotting the computed synthetic spectra with high and low abundances for each molecule. Spectral features of H$_{2}$O and CO$_{2}$ affect the spectrum at all wavelengths. CO has absorption features at 1.6 --1.7, 2.3--2.5, and 5.0--5.5 $\mu$m, while CH$_{4}$ features can be seen at 1.7, 2.2--2.5, 3--4, and 7--9 $\mu$m. A striking feature of the best-fitting spectrum is a deep IR absorption by CO$_{2}$ at 9--24 $\mu$m, which is rather different from the fitting by other studies \citep[MS09;][]{for07,gri08,cha08}, who concluded that the features at the longer wavelengths of IRS are caused mainly by H$_{2}$O, of which absorption is also dominant at the shorter wavelengths of 5--9 $\mu$m.  Although the low CO$_{2}$ hypothesis seemed to give a good reproduction of the long-wave $Spitzer$ data, it fails to reproduce the shorter wavelengths covered by $HST$. As seen here, one of the benefits of our approach is that the use of a broad range of IR wavelengths allows us to break some of the degeneracies inherent in modelling a small number of data points (cf. Section 5.4).

In this study, the high CO$_{2}$ abundance causes a sharp drop at 9 $\mu$m and then a flat and featureless spectrum between 9--24 $\mu$m. H$_{2}$O still contributes to the spectrum at 5--9 $\mu$m and $>$20 $\mu$m as well as a small feature from a vibrational transition at 6--6.5 $\mu$m, which is one of the H$_{2}$O features highlighted by \citet{gri08}. In particular, the spectrum fits the planet-star flux ratio for the $Spitzer$ IRAC 3.6 $\mu$m channel, a feature which has been previously explained by the large 3.25 $\mu$m emission of CH$_{4}$ in non-local thermodynamic equilibrium (NLTE) conditions \citep{swa10,tha10}. However, any additional NLTE calculation is not required in our model to produce the data at 3.6 $\mu$m adequately. For the $HST$/NICMOS channels between 1.45--2.5 $\mu$m, we fit to the measurements using a high amount of CO$_{2}$ as suggested in MS09 ($\sim$7$\times$10$^{-4}$) rather than in S09 whose model suggested low CO$_{2}$ mixing ratio (10$^{-7}$--10$^{-6}$). This discrepancy is not yet fully resolved, but, as \citet{sha11} stated, this may come from the difference in the forward models used for each study. In our spectrum, the channels at 1.584, 1.869, 2.159--2.216 $\mu$m, however, are underestimated by the high CO$_{2}$ abundance, leading to a decrease in the quality of the fit to the data. The sharp edge present in our synthetic spectrum at 9 $\mu$m, where a sudden CO$_{2}$ line weakening is shown \citep{tas03}, is caused by a strong CO$_{2}$ absorption. Its absence from the measured exoplanet spectrum is yet to be understood but may be related to the limited number of gases (H$_{2}$O, CO$_{2}$, CO, and CH$_{4}$) and the absence of absorbing aerosols considered in our modelling.

\section{Retrieval of atmospheric properties}
On the basis of the best-fitting spectrum presented in the previous section, we here validate the retrieval method by analysing the temperature contribution functions and the functional derivatives for the molecular abundances. These sensitivity analyses then allow plausible interpretations for temperature and composition structure of the dayside of HD 189733b. 

\subsection{Contribution functions}

\begin{figure*}
\centering
\includegraphics[width=14.5cm]{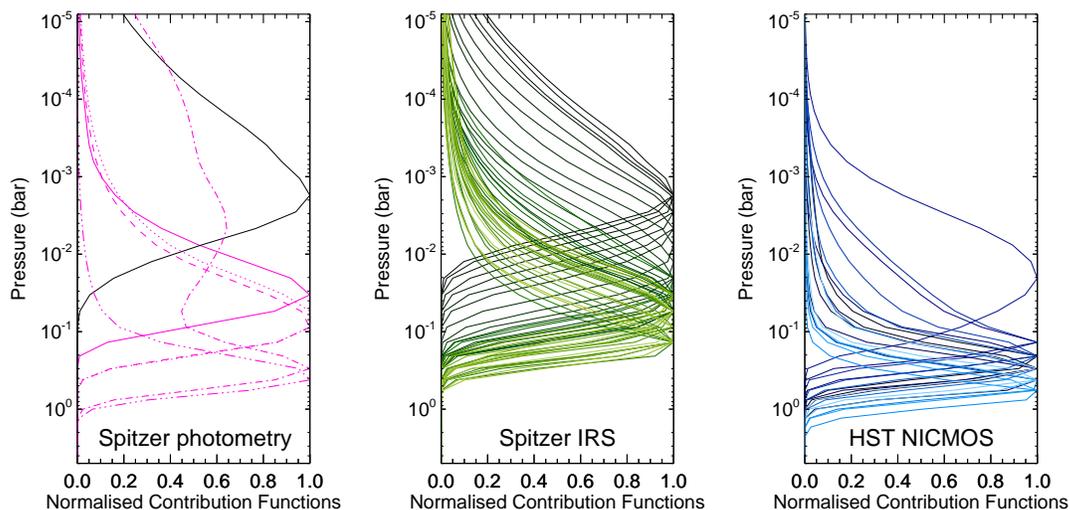}
\caption{Contribution functions for the $Spitzer$ broadband photometry (left), IRS spectroscopy (middle), and the $HST$/NICMOS spectrophotometry (right) channels. For the $Spitzer$ photometry channels, each line pattern means MIPS 24 $\mu$m (solid), IRS 16 $\mu$m (solid-black), IRAC 8.0 $\mu$m (dotted), 5.8 $\mu$m (dashed), 4.5 $\mu$m (dot-dashed), and 3.6 $\mu$m (triple dot-dashed). For the $Spitzer$ IRS and the $HST$/NICMOS channels, the brighter colours denote the channels at the shorter wavelengths. For all cases, emission from the lower atmosphere tends to dominate the shorter wavelength channels.}
\label{f3}
\end{figure*}

\begin{figure*}
\centering
\includegraphics[width=10cm]{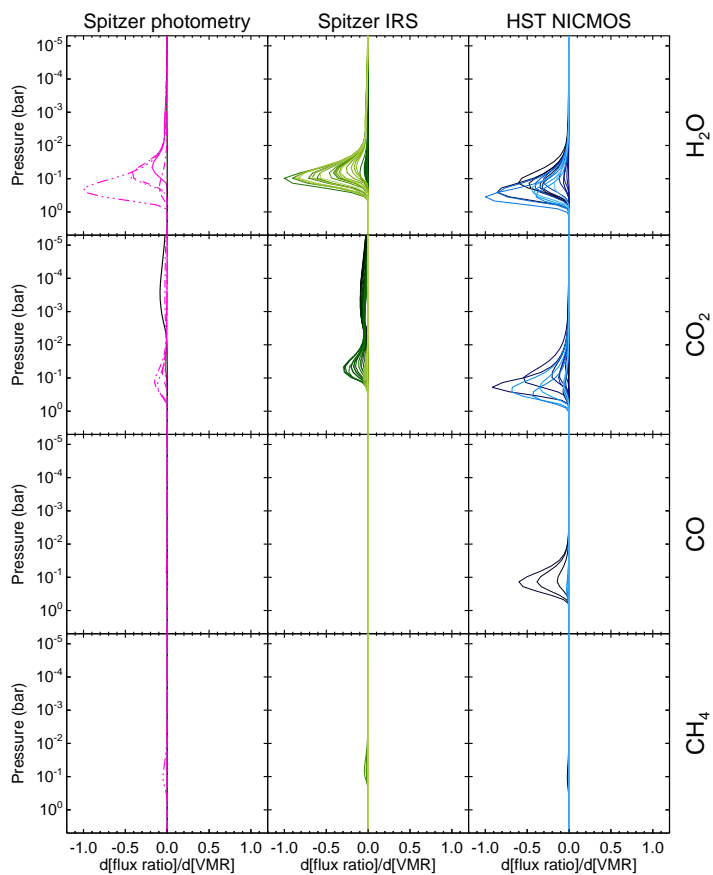}
\caption{Normalised functional derivatives of molecules in $Spitzer$ and $HST$ channels. The applied colours are the same as those described in Fig.~\ref{f3}. Each row shows the vertical sensitivity of radiance with respect to the abundances H$_{2}$O, CO$_{2}$, CO, and CH$_{4}$ (from top to bottom). This figure shows the pressure levels at which molecular abundance can be retrieved (row direction) from which channel (column direction).}
\label{f4}
\end{figure*}

The radiance measured in each channel originates from different pressure levels within the atmosphere. This is generally described using a contribution function, which is the product of the Planck function at the local temperature in the atmosphere and the transmission weighting function, which describes the rate of change of atmospheric transmission with height. The contribution functions are dependent on the best-fitting profiles for the atmospheric structure and abundances. Thus the calculated contribution functions indicate the pressure levels at which thermal emission from the atmosphere contributes most to the radiance observed in each channel. Fig.~\ref{f3} shows the contribution functions for all channels of the $Spitzer$ IRS, IRAC, and MIPS, and $HST$/NICMOS.

The six $Spitzer$ broadband photometry channels in the range 3.6--24 $\mu$m have broadly-distributed contribution functions whose peak pressures range from 2 to 300 mbar. The contribution functions for the IRAC channels (3.6, 4.5, 5.8, and 8 $\mu$m) and MIPS (24 $\mu$m) are located in the deeper atmosphere and provide strong constraints for the temperature between 30--300 mbar. The contribution function of the IRAC 4.5 $\mu$m channel has a second peak at high altitude (3--5 mbar), being close to the peak of the IRS 16 $\mu$m channel at 2 mbar. Hence, the temperature at pressures as low as 2 mbar can be retrieved from the IRAC channels, but only if the temperature of the deep atmosphere is well constrained from the other measurements. The 47 $Spitzer$ IRS spectroscopy channels between 5 and 14.5 $\mu$m have closely-spaced overlapping contribution functions, with peaks ranging from 2 to 150 mbar. The radiance from the lower atmosphere ($\sim$100 mbar) contributes to the channels in the range 5--9 $\mu$m, whereas radiance from the upper atmosphere ($\gtrsim$10 mbar) contributes longward of 9 $\mu$m. Unlike the MIR and FIR channels on $Spitzer$, the $HST$/NICMOS NIR 1.45--2.5 $\mu$m channels can measure the emission from the deeper atmosphere at 20--600 mbar and their contribution functions are partly distributed over the pressure levels that are not covered by the $Spitzer$ measurements. 

In summary, the contribution functions show that the $P$-$T$ profile over a broad range of pressures from 2 to 600 mbar can be constrained by retrieving from the $Spitzer$ and $HST$ measurements together. On the other hand, studies that focus on a single narrow wavelength range are sensitive to only a narrow altitude range, with potential degeneracies between temperatures at high and low altitudes. Thus, by considering all available datasets simultaneously, we can provide stronger constraints on the retrieved thermal properties than in previous studies.

\subsection{Functional derivatives}

The functional derivatives are defined to be the partial derivatives of the radiance (or any spectral output in the forward model) with respect to any given atmospheric parameters. By calculating the functional derivatives for molecular abundances, we can understand which measurements are sensitive to the molecular abundance and at which pressure levels. Fig.~\ref{f4} shows the functional derivatives for the 4 molecules considered in this study. For ease of comparison, these are normalised to the peak of the functional derivatives for each measurement set. Using Fig.~\ref{f4} the sensitivity can be interpreted in two directions: Each row indicates which pressure levels show high sensitivity to a given molecule in each channel, and each column indicates how molecular sensitivity is distributed through pressure levels in a given set of measurements. In all cases, the derivatives are negative because an increasing abundance increases the atmospheric absorption (in the absence of a thermal inversion on this planet) and decreases the ratio of the disk-averaged planetary flux with respect to the host star.

High sensitivity to the abundance of H$_{2}$O is seen in all channels, with the exception of the 1.93--2.16 $\mu$m channels of $HST$/NICMOS, 4.5 $\mu$m channel of $Spitzer$ IRAC, 9--14.5 $\mu$m and 16 $\mu$m channels of the $Spitzer$ IRS, in which regions the modelled radiance is dominated by CO$_{2}$ absorption, as explained below. The H$_{2}$O functional derivatives all peak in the $\sim$100--500 mbar region, showing that the measurements can only constrain the H$_{2}$O abundance in this altitude range. In contrast, the CO$_{2}$ functional derivatives are divided into two separate pressure levels (peaks between $\sim$0.1 bar and 0.1--1 mbar) as are clearly shown in the second row of Fig.~\ref{f4}. The $HST$/NICMOS channels between 1.584--1.698 and 1.929--2.216 $\mu$m are only sensitive to the CO$_{2}$ abundance at altitudes below 100 mbar whereas the $Spitzer$ broadband photometry and IRS spectroscopy are sensitive to the CO$_{2}$ in the 0.1--1 mbar region as well. Despite this sensitivity to a range of altitudes for CO$_{2}$, a combination of these channels, however, may not determine CO$_{2}$ abundance at both levels due to the small sensitivity in the upper atmosphere (cf. Section 5.4). CO can only be detected in three $HST$/NICMOS channels in the range 2.33--2.45 $\mu$m at 100 mbar, and these channels have been extensively used to constrain the CO mixing ratio (S09 and MS09). The sensitivities for CH$_{4}$ are so marginal, as seen in the fourth row, that constraining its abundance may not be possible from the given measurements. 

In summary, the functional derivatives indicate the altitudes and channels showing sensitivity to each particular molecule. The sensitivities to the molecules are mostly clustered in the lower atmosphere ($\sim$100 mbar), and, in particular, CO$_{2}$ shows an additional peak at $\sim$1 mbar. Therefore the molecular abundances can only be constrained in the deep atmosphere (troposphere), with less sensitivity to the upper atmosphere. Finally, these functional derivatives show that the spectrum is sensitive to the H$_{2}$O abundance over a broad range in most of the measured channels, implying an inherent degeneracy between temperatures and H$_{2}$O in these observations.

\begin{figure}
\includegraphics[width=8.7cm]{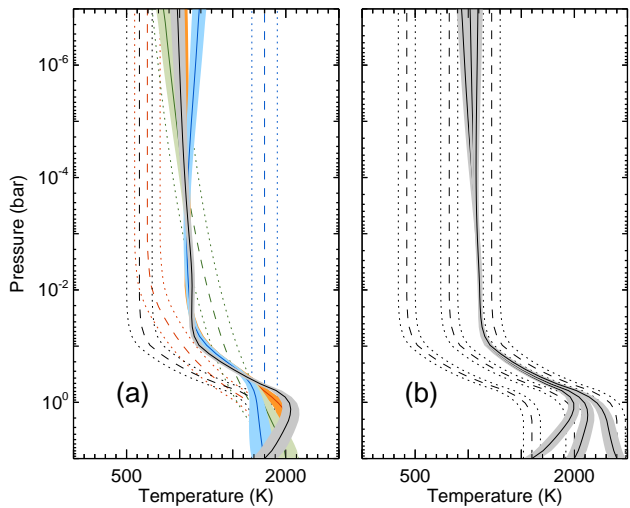}
\caption{(a) Retrieved $P$-$T$ profiles of the dayside of HD 189733b from a range of diverse $a$ $priori$ profiles. The different lines show: $a$ $priori$ $P$-$T$ profiles (dashed), retrieved $P$-$T$ profiles (solid) and their errors (dot-dashed and dotted). This plot shows the pressure range over which the $P$-$T$ profile is retrievable from these measurements. (b) Retrieved $P$-$T$ profiles from the same $a$ $priori$ profile shape, but offset from each other with baseline temperatures at 1600 K, 2000 K, and 2400 K at 10 bar.}
\label{f5}
\end{figure}

\subsection{Retrieval of $P$-$T$ profile}

\subsubsection{The best-fitting $P$-$T$ profile}
Previously $P$-$T$ profiles of HD 189733b have been estimated based on theoretical models or parametric retrievals to generate many thousands of model spectra to compare with observations. However, this forward modelling approach does not explicitly solve the inverse problem during the constraining process and thus it is unclear if the solutions are biased more towards theoretical expectations than being driven by the measurements themselves. For this reason, we derive temperatures using several different $a$ $priori$ profiles to show that the retrieved temperature converges to a reproducible profile in the altitude range covered by the contribution functions. Because a biased $P$-$T$ profile would vary with the $a$ $priori$ assumptions, we look for an appropriate $a$ $priori$ error and its vertical shape in order that the same $P$-$T$ profile can be retrieve irrespective of the shape of the $a$ $priori$ profile. Fig.~\ref{f5}(a) presents the retrieved $P$-$T$ profile and its error for a range of selected temperature $a$ $priori$. For all cases, the $P$-$T$ profiles share a common shape between 0.1 mbar and 1 bar, demonstrating the validity of the temperature retrieval over this range. It is shown that even with the simplest possible assumption such as an isothermal temperature throughout the atmosphere [blue line in Fig.~\ref{f5}(a)], the measurements still produce a similar thermal profile as the other retrievals. As a further test, we take an $a$ $priori$ structure from the retrieved profile in the previous step, offset it by a temperature 400 K, and repeat the temperature retrieval again. Fig.~\ref{f5}(b) shows that the $P$-$T$ profile is still sufficiently constrained by the measurements, even if there are large shifts at levels not covered by the contribution functions, where the solutions relax back to their different $a$ $priori$s. We conclude here that the available dayside emission spectra can constrain an unbiased atmospheric structure over the levels probed by the contribution functions.

As a result, we find that the temperature decreases adiabatically from 1900 K at 600 mbar to $\sim$1200 K at 100 mbar, then becomes isothermal up to the upper atmosphere ($\sim$1 mbar). These adiabatic and isothermal layers in the thermal structure are dominant features of heat transfer by convection and radiative cooling, respectively. In comparison S09 claimed a decreasing temperature layer between 0.01 and 1 bar to model the $HST$/NICMOS measurements, making an adiabatic layer $\sim$10 times thicker than our estimation, and theoretical models also considered adiabatic layers in the troposphere \citep{for06,bur08,sho08}. The isothermal temperature ($\sim$1100 K) of the dayside hemisphere at pressure levels above the troposphere (100 mbar) is possibly maintained by efficient energy re-distribution throughout the whole planet system as suggested by the observations in the $Spitzer$ IRAC channels \citep{knu07,cha08}. Various circulation models \citep{sho08,dob08} predict that the heat transports between the dayside and nightside lead to the isothermal structure at mid and low pressure ($<$100 mbar) in the dayside atmosphere.

In summary, the broad wavelength range of available measurements has provided a strong constraint on the vertical $P$-$T$ profile (an adiabatic troposphere and isothermal stratosphere) without excessive sensitivity to the $a$ $priori$ assumptions, and without reliance on theoretical modelling or parameterised profiles.

\subsubsection{Temperature degeneracy}

\begin{figure}
\includegraphics[width=8.6cm]{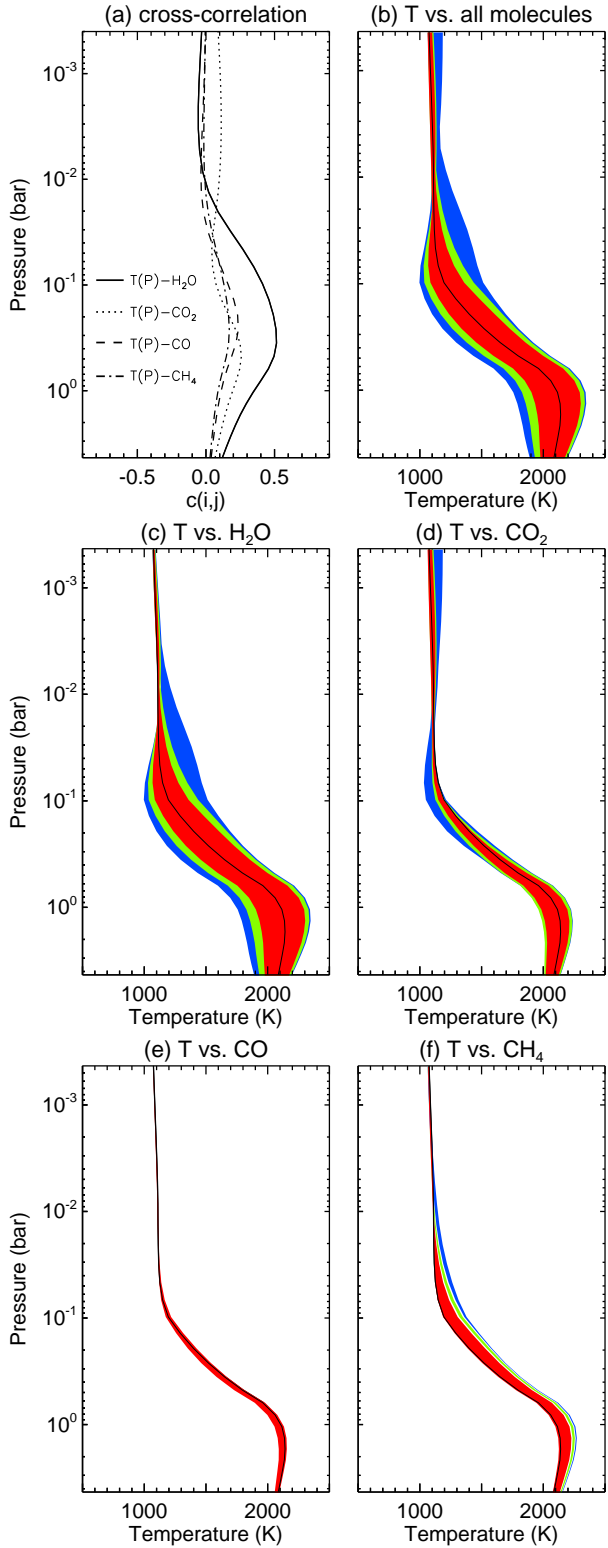}
\caption{(a) The vertical structure of the cross-correlation functions. The high correlation with H$_{2}$O and CO$_{2}$ dominates the degeneracy of temperature at most pressures. (b-f) The retrieved $P$-$T$ profiles of the dayside HD 189733b with various mixing ratios for each molecule are presented with $\Delta\chi^{2}/N$ ranges. Each colour demonstrates $\Delta\chi^{2}/N$ $<$0.5 (red), $<$1.0 (green), and $<$2.0 (blue), respectively.  As a result, the large uncertainties of $P$-$T$ profile are broadly distributed between the deep and middle atmosphere. On the other hand, narrow uncertainty at 2 mbar implies an isothermal structure in the upper atmosphere of the dayside of HD 189733b.}
\label{f6}
\end{figure}

In general, the best-fitting $P$-$T$ profile from the retrieval is a non-unique solution. This means that there is potentially a large family of solutions for the $P$-$T$ profile that could fit the measured spectrum equally well within the same error range. In many retrievals, there may also exist cross-correlation with other fitted quantities, indicated by covariances that are larger than zero. The covariance matrix is advantageous because the matrix is computed as a part of the retrieval process and can be used in assessing the degeneracies between variables at the end of retrieval. 

For the dayside spectrum of HD 189733b, the degeneracy between the molecular abundances and the $P$-$T$ profile is thought to be significant at some pressure levels and bandpasses (S09 and MS09), however, a detailed analysis of these correlations has not previously been presented. Here, we examined the cross-correlation functions $\it{c(i, j)}$ (i.e. the off-diagonal elements of the correlation matrix) that determine the degree of degeneracy between elements $i$ and $j$ of the state vector. By definition, $|\it{c(i, j)}|$=1.0 represents a perfect correlation between variable $i$ and $j$, and we consider that $|\it{c}|$=0.5 is a limitation for quasi-independent retrievals of different variables. The profiles in Fig.~\ref{f6}(a) show the vertical structure of cross-correlation functions between the molecular abundances and the temperature at different pressure levels. It can be seen that the $P$-$T$ profile is significantly correlated with molecular abundances at some levels, being most correlated with the H$_{2}$O abundance between 200--400 mbar and with CO$_{2}$ for altitudes above the 30 mbar pressure level.

To determine the degeneracy of the temperature profile from Fig.~\ref{f6}(a), we performed multiple retrievals where individual molecular abundances were fixed at a particular value and all other variables retrieved. We then assessed the weighted mean squared error, the $\chi^{2}$ over the number of measurements ($N$) of the solution as a function of the fixed molecular abundance, compared with the minimum $\chi^{2}/N$  ($\chi^{2}_{min}/N$) of the best-fitting retrieval, and determined the range of temperatures for which the change $\Delta\chi^{2}/N=\chi^{2}/N-\chi^{2}_{min}/N$ was less than 0.5, 1.0, and 2.0, respectively.

Fig.~\ref{f6}(b) shows the ranges in the retrieved $P$-$T$ profile for different values of $\Delta\chi^{2}/N$ due to the molecular degeneracy. At each pressure level, allowed temperature ranges are determined by varying the molecular abundance with which temperature is most strongly correlated. The temperature uncertainties for $\Delta\chi^{2}/N<1.0$ calculated in this way are 540 K at 600 mbar, 420 K at 100 mbar, and 100 K at 2 mbar. As described on the basis of cross-correlation, the temperature uncertainties at altitudes below 10 mbar are caused mainly due to the degeneracy with the H$_{2}$O abundance and those at altitudes above 10 mbar are degenerate with the CO$_{2}$ abundance. On the other hand, the degeneracies between temperature and the abundances for CO and CH$_{4}$ are small. Therefore it can be seen that although some uncertainty in the deeper $P$-$T$ profile exists due to molecular degeneracy, the small degree of degeneracy at lower pressures supports our conclusion that HD 189733b has an isothermal structure at the upper atmosphere.

\begin{table*}
\begin{center}
 \caption{Estimated and retrieved mixing ratios and C/O in the dayside of HD 189733b.}
\label{t1}
\begin{tabular}{|c|c|c|c|c|c|c|}
\hline
& H$_{2}$O (10$^{-4}$) & CO$_{2}$ (10$^{-4}$) & CO (10$^{-4}$) & CH$_{4}$ (10$^{-4}$) & C/O & Data Source\\
\hline
\multirow{2}{*}{Line et al. 2010} & 6--13 & 0.0047--0.016 & 2--9 & 0.0026--6.758 &  & Thermochemistry\\
 & $\sim$6.36 & 0.004--$\sim$0.1 & $\sim$8.4 & $\sim$0.4 & & Photochemistry\\
\hline
Swain et al. 2009 & 0.1--1 & 0.001--0.01 & 1--3 & $<$0.001 & 0.5--1 & $HST$/NICMOS\\
\hline
\multirow{3}{*}{Madhusudhan \& Seager 2009} & 0.01--1000 & ... & ... & $<$100 & ... & $Spitzer$ IRS\\
 & 0.1--10 & 0.007--0.7 & ... & $<$0.02 & 0.007--1 & $Spitzer$ photometry\\
 & $\sim$1 & $\sim$7 & 2--200 & $<$0.06 & 0.5--1 & $HST$/NICMOS\\
\hline
 & \multicolumn{6}{|c|}{Best-fitting value} \\ 
 & 6.7 & 20 & 29 & 0.0019 & 0.65 & All measurements \\ \cline{2-7}
This study & \multicolumn{6}{|c|}{Possible fit range} \\ 
\multirow{3}{*}{} & 0.9--50 & 3--150 & ... & $<$0.4 & 0.45--1 & $\Delta\chi^{2}/N$ $<$ 0.5 \\
 & 0.3--100 & 1.5--300 & ... & $<$1 & 0.30--1 & $\Delta\chi^{2}/N$ $<$ 1.0\\
 & 0.03--400 & $<$2000 & ... & $<$3 & 0.15--1 & $\Delta\chi^{2}/N$ $<$ 2.0\\
\hline
\end{tabular}
\end{center}
\end{table*}

\subsection{Retrieval of molecular abundances}
Despite the molecular degeneracies described above, we can use the dayside emission spectra to place crude estimates on the abundances in addition to the vertical temperature structure.  Each gas is assumed to be well-mixed with a constant mole fraction with height. However, we know that the functional derivatives are only sensitive to a limited pressure range and thus our retrieved abundances represent mean concentrations only at the pressure levels which show the highest sensitivity (Fig.~\ref{f4}). The best-fitting molecular abundances are determined from the retrieval accompanying the best-fitting $P$-$T$ profile given in Section 5.3 and we find that those are 6.7$\times$10$^{-4}$ (H$_{2}$O), 2.0$\times$10$^{-3}$ (CO$_{2}$), 2.9$\times$10$^{-3}$ (CO), and 1.9$\times$10$^{-7}$ (CH$_{4}$), respectively. Due to its tiny contribution to the spectrum, the best estimation of CH$_{4}$ should be considered as an upper limit only, and the CO abundance is poorly constrained by the available data (see below). Additionally, the C/O ratio of HD 189733b (0.65) is found to be close to the solar value ($\sim$0.6).

\begin{figure}
\hspace{-0.9cm}
\includegraphics[width=9.5cm]{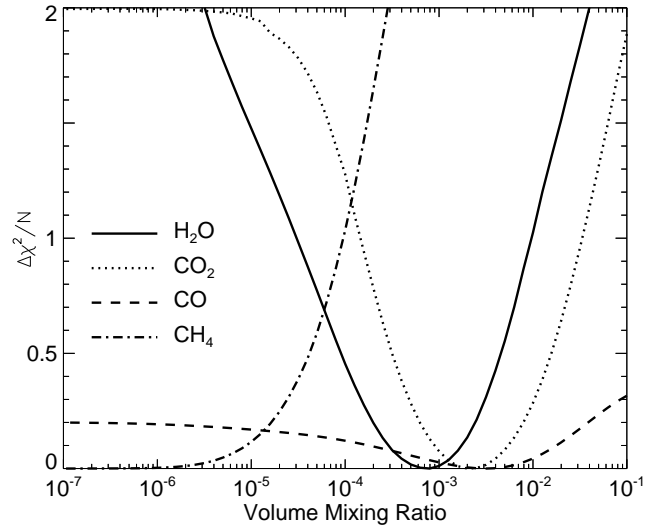}
\caption{The degeneracy ranges of the molecular mixing ratios for H$_{2}$O, CO$_{2}$, CO, and CH$_{4}$. Each line shows resultant $\chi^{2}/N$ with respect to given abundances. The $\Delta\chi^{2}/N$ constrains the uncertainties of the molecular abundances which are distributed around the best-fitting abundance of H$_{2}$O (6.7$\times$10$^{-4}$), CO$_{2}$ (2.0$\times$10$^{-3}$), CO (2.9$\times$10$^{-3}$), and CH$_{4}$(1.9$\times$10$^{-7}$), respectively.  Lower bounds of CO and CH$_{4}$ uncertainties are unconstrained because of their small contribution to the spectrum.}
\label{f7}
\end{figure}

By looking at the covariance of our retrieved solution we find that the cross-correlation between the different molecular abundances is rather small ($c < 0.1$) for all the molecular combinations. However, there remains some degeneracy due to the main correlation with temperature.  To quantify this we calculated $\Delta\chi^{2}/N$ for each gas by changing the abundance of a molecule and retrieving all other parameters including temperature. The variation of $\Delta\chi^{2}/N$ with abundance for all four gases is shown in Fig.~\ref{f7}. We find that the following abundance ranges have $\Delta\chi^{2}/N$ $<$ 0.5: (9--500)$\times$10$^{-5}$ for H$_{2}$O, (3--150)$\times$10$^{-4}$ for CO$_{2}$, and $<$4$\times$10$^{-5}$ for CH$_{4}$, respectively. The error ranges for the abundance of CO were so broad that we cannot provide meaningful estimates from the dayside spectra alone, hence they are absent from Tab.~\ref{t1}.   The retrieved abundances of the other species for $\Delta\chi^{2}/N$ $<$ 1.0 and 2.0 are also shown in Tab.~\ref{t1} and are compared with the expectations of a thermo- and photo-chemistry model \citep{lin10}, and with previous forward model studies (S09 and MS09). As we found using the $F$--test in Section 4.1, the data only constrain the upper bound of the mixing ratio for CH$_{4}$ and do not offer useful constraints on the abundance of CO because the disk-averaged planet-star flux ratio shows only a small sensitivity to their abundances in Fig.~\ref{f4} and their contributions to the emission spectrum appear in only few bandpasses. Consequently, variation of these two gases leads to a tiny change in $\Delta\chi^{2}/N$ where the abundances are lower than the best-fitting estimation. We conclude that it is not possible to constrain the abundances of CO and CH$_{4}$ based on these datasets alone as previous studies have attempted.

All channels of $Spitzer$ and $HST$ are capable of constraining the H$_{2}$O mixing ratio as presented in the functional derivatives, and the constrained H$_{2}$O is consistent with other estimations. For CO, the strongest constraints are taken from the $HST$/NICMOS data (MS09) and we find that the dayside emission spectra can offer no meaningful constraints on the abundance of this species. A relatively small abundance of CH$_{4}$ has been estimated by previous studies in Table 1, except for the large abundance (10$^{-2}$) derived from the $Spitzer$ IRS spectroscopy at 7.6 $\mu$m (MS09), which is not consistent with this study. The derived abundance of CO$_{2}$ (3--150$\times$10$^{-4}$) is 2--3 orders of magnitude larger than previous studies by \citet{lin10} ($\sim$10$^{-7}$--$\sim$10$^{-5}$), S09 (10$^{-7}$--10$^{-6}$), and MS09 (using $Spitzer$ photometry, 7--700$\times$10$^{-7}$) for all $\Delta\chi^{2}/N$ ranges and thus our retrieved CO$_{2}$ abundance is only consistent with the $HST$/NICMOS estimation by MS09. Indeed, it is the $HST$/NICMOS channels that constrain our retrieved CO$_2$ abundance to be larger than previous studies ($\sim$10$^{-3}$), whereas the $Spitzer$/IRAC channels at 4.5 and 16 $\mu$m could be fitted with either low or high abundances of this molecule.

\begin{figure}
\hspace{-0.3cm}
\includegraphics[width=8.9cm]{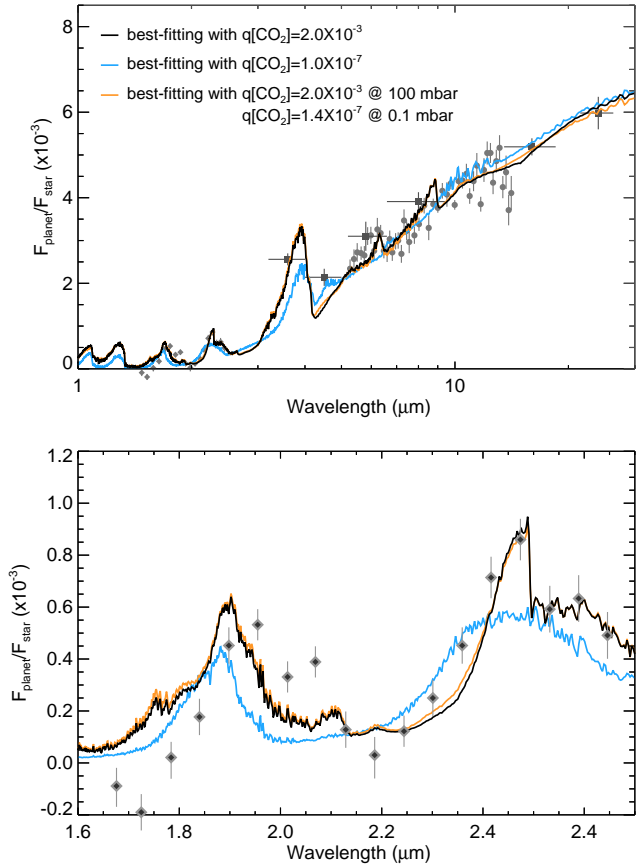}
\caption{Synthetic emission spectra of the dayside of HD 189733b with different CO$_{2}$ abundances. The three lines show the spectrum with a well-mixed abundance of 2.0$\times$10$^{-3}$ (black); well-mixed at 1$\times$10$^{-7}$ (cyan) and the case with CO$_2$ varying with altitude from 2.0$\times$10$^{-3}$ at 100 mbar to 1.4$\times$10$^{-7}$ at 0.1 mbar (orange). The latter case is almost indistinguishable from the first scenario.}
\label{f8}
\end{figure}

To test the sensitivity to CO$_{2}$ in the upper atmosphere, we compared the best-fitting spectra to test cases with a range of upper atmospheric abundances of this molecule (i.e. allowing the abundance to vary with altitude).  The CO$_{2}$ mixing ratio was fixed at a certain value up to a pre-defined pressure level, and then declined linearly into the upper atmosphere with a slope determined by a fractional scale height (the ratio of the scale height of the gas to the scale height of the atmosphere). This allows us to demonstrate the effect of different mixing ratios with height, but avoiding the need to introduce a complete vertical profile of CO$_{2}$ (and hence a large number of additional parameters in our state vector). In Fig.~\ref{f8}, the retrieval with the new CO$_{2}$ vertical profile that decreases from 2.0$\times$10$^{-3}$ at 100 mbar to 1.4$\times$10$^{-7}$ at 0.1 mbar, produces a very similar spectrum to the well-mixed high-CO$_{2}$ case, leading to an insignificant improvement in terms of the $\chi^{2}/N$. This is because the CO$_{2}$ sensitivity at the $Spitzer$ bandpasses is too low to adequately constrain its abundance in the upper atmosphere (0.1--1 mbar), whereas the strong constraints of the $HST$/NICMOS and $Spitzer$ IRS channels between 9--16 $\mu$m allow us to constrain the abundance in the lower atmosphere ($\sim$100 mbar). In contrast, a small abundance of CO$_{2}$ (10$^{-7}$) with a well-mixed vertical profile is unable to reproduce the measured spectrum.   Our results suggest that a large amount of CO$_{2}$ ($\sim$10$^{-3}$) exists at $\sim$100 mbar but that the abundance at 0.1--1 mbar is unconstrained by current measurements.

\subsection{Additional Sources of Degeneracy}
In the sections so far, we have derived conclusions about the atmospheric temperatures and composition based on a series of underlying assumptions, which have not been quantified by previous authors.   Here we perform additional degeneracy tests to quantify the uncertainties on temperature and composition caused by the uncertainty of the amount of H$_{2}$ and He and the radius ratio between planet and star ($R_{p}/R_{\ast}$). The abundances of H$_{2}$ and He are generally considered to be more or less the same as in the parent star.  However, the helium abundance in the giant planets of our own solar system deviates considerably from solar, and this could provide an additional source of error in the retrieval of the vertical temperature structure.   However, we find that the change in mole fraction of H$_{2}$ (assumed to be 1-q[He]) between 0.8--1.0 gives a negligible effect on the retrievals of temperature and composition from the dayside spectra at all pressure levels, suggesting that the emission spectrum of HD 189733b (which is dominated by the features of H$_2$O) is not particularly sensitive to variations of H$_{2}$ and He compared to the host star.

\begin{figure}
\includegraphics[width=9cm]{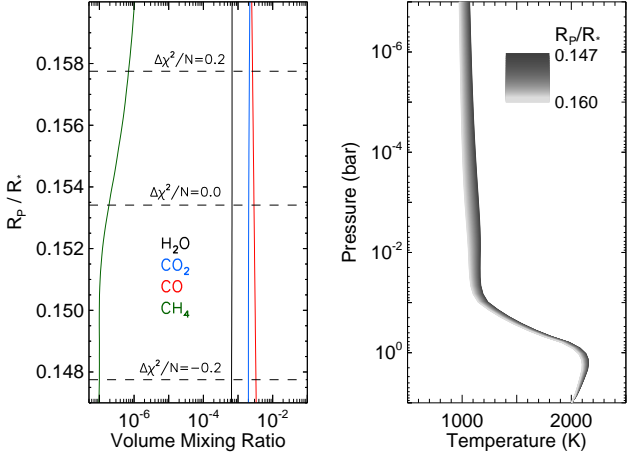}
\caption{Uncertainty of composition (left) and temperature (right) being induced by the variation of planet-star radius ratio ($R_{p}/R_{\ast}$). Due to $\sim$10 per cent increment in $R_{p}/R_{\ast}$, the temperature profile shifts towards lower $\sim$100 K whereas the fitting quality, $\Delta\chi^{2}/N$ increases by 0.5.}
\label{f9}
\end{figure}

On the other hand, variations of the planet-star radius ratio, which determines the bottom level of the model atmosphere, has a significant effect on the flux ratio.  If we consider the effect of varying the radius ratio over the range 0.147$<$$R_{p}/R_{\ast}$$<$0.160 \citep{ago10}, then we find that the best-fitting molecular abundances vary over the ranges: H$_{2}$O (6.5--6.7$\times$10$^{-4}$), CO$_{2}$ (2.0--2.2$\times$10$^{-3}$), CO (2.5--3.3$\times$10$^{-3}$), and CH$_{4}$ (1--10$\times$10$^{-7}$), respectively (Fig.~\ref{f9}). The figure indicates that the temperature profile over the pressure range becomes $\sim$100 K cooler when $R_{p}/R_{\ast}$ is enhanced by 0.013 ($\sim$10 per cent increase in the radius of the planet). As an increase in $R_{p}/R_{\ast}$ enhances the flux ratio across all wavelengths, the model tends to select a correspondingly low temperature profile that decreases the flux ratio.

Finally, as we described in Section 3, we have omitted the scattering and absorption processes due to the present of clouds and hazes in this hot Jupiter atmosphere, choosing to focus on a clear atmosphere for this initial exercise.  Given the present quality of the dayside emission spectra, the more complex model with scattering hazes is not needed to reproduce the data (although hazes are clearly needed to reproduce transmission spectra of the terminator region), and would lead to a significant rise in the number of model parameters.  However, the omission of aerosols could lead to significant changes to our derived temperatures and molecular abundances,  and our future work will aim to quantify these uncertainties to further explore the atmospheres of hot Jupiters.

\section{Discussion and Conclusions}

\begin{figure}
\hspace{-0.2cm}
\includegraphics[width=9cm]{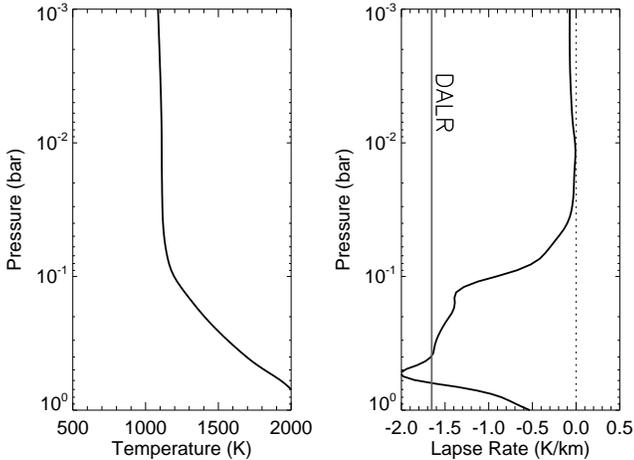}
\caption{Disk-averaged dayside lapse rate of HD 189733b (right) from the derivative of the $P$-$T$ profile (left). The given pressures correspond to the levels covered by the contribution functions. We also display the dry adiabatic lapse rate (DALR) in grey, assumed to be a constant over this pressure range.}
\label{f10}
\end{figure}

The retrieved thermal structure is a useful tool for further understanding of dynamic processes in the atmosphere. As a description of atmospheric stability, the lapse rate ($\Gamma=dT/dz$, the rate of change of temperature with respect to the height) is shown in Fig.~\ref{f10}.  An increase in the lapse rate indicates that there is an adiabatic layer beneath the $\sim$100 mbar level ($\sim$1400 km above from the level of 10 bar), and that the lapse rate tends to zero for pressure levels above $\sim$30 mbar ($\sim$2100 km) where the temperature profile tends to an isotherm. Moreover, the measured lapse rate is similar to the dry adiabatic lapse rate (DALR) below the 300 mbar pressure level, and is overlain by a sub-adiabatic layer extending up to 30 mbar, similar to the atmospheric thermal structures seen in the upper tropospheres of the giants planets of our solar system.

The NEMESIS algorithm uses two novel tools for an efficient retrieval of the atmospheric properties: a non-linear optimal estimation scheme and a fast forward model taking advantage of the correlated-$k$ approximation and $k$-distribution tables. This rapid retrieval method allows us to infer general atmospheric properties from small datasets. The optimal estimation retrievals permit an extension of previous results and the formal quantification of errors and uncertainties. The functional derivatives, which are analytically computed, enable us to trace the vertical contribution of the different input constituents in each channel, permitting us to move beyond the single best-fitting approach and show the relationship between the features of the data and physical values. We have shown that the current set of observations of the dayside spectrum of HD 189733b are enough to constrain the thermal structure at some pressure levels and the mixing ratios for H$_{2}$O and CO$_{2}$, albeit with large uncertainties due to molecular degeneracies and the limited number of spectral measurements. On the other hand, we do not find statistically-significant evidence for CO and CH$_4$ in the dayside emission spectra alone, and the present observational set of HD 189733b only allows us to place upper limits on the abundances of these species. Furthermore, the ability to calculate the cross-correlation function allows us to assess the degeneracies between the various modelled parameters in our state vector.

With the retrieval method presented here, we derive three major findings from the measurements of the dayside of HD 189733b. Firstly, we have produced the first retrieval (i.e. solution to the inverse problem) of the vertical temperature structure of the dayside atmosphere of HD 189733b. The retrieved profile shows that the constraints in the NIR and MIR lead to an adiabatically decreasing temperature between 0.1--1 bar. In addition, strong constraints from the $Spitzer$ IRS measurements in the 9--16 $\mu$m range suggest that the temperature structure in the mid and upper atmosphere (1--100 mbar) seems to be isothermal with a uniform temperature ($\sim$1100 K). One explanation for the isothermal layer is that super-rotating jets blowing from the night- to dayside play an critical role of an efficient energy re-distribution over the whole planet and may be responsible for maintaining the high temperature in the upper atmosphere \citep{knu07,sho08}.

Secondly, the functional derivatives for the molecules show that the measurements at NIR (1.6--1.7 and 1.9--2.2 $\mu$m), MIR (4.5 $\mu$m), and FIR (9--16 $\mu$m) spectral ranges constrain the CO$_{2}$ abundance at $\sim$100 mbar but not in the upper atmosphere. Its abundance in the lower atmosphere is responsible for the spectral flatness at 9--16 $\mu$m, where previous studies assumed H$_{2}$O to be the dominant constituent. We determine the vertical sensitivity of CO$_{2}$ abundance by comparing the retrievals with abundant (2.0$\times$10$^{-3}$) and scarce CO$_{2}$ (10$^{-7}$), and height-dependant CO$_{2}$ profile. The functional derivatives for CO$_{2}$ show low sensitivity at 0.1--1 mbar, which implies that the abundance of CO$_{2}$ in the upper atmosphere remains underconstrained by existing data. 

Thirdly, we have quantified the degeneracies between the atmospheric properties using the cross-correlation functions. The correlations between molecular abundances and temperature exhibit substantial degeneracies, in particular, between temperature and H$_{2}$O abundance at 300 mbar. We determine the temperature uncertainty at different pressure levels by calculating the statistics of the retrievals with the variation of each molecular mixing ratio. In the same way, the molecular abundances are also constrained by demonstrating that the uncertainty on retrieved parameters, based on the available measurements and the degeneracy with the temperature profile is considerably larger than previous studies have suggested. Therefore, additional data is clearly required to break the degeneracy between temperature and composition, particularly H$_{2}$O. As long as the number of retrieval variables for the molecular abundance are kept to a small number, the cross-correlation functions are not significant between molecules. If we were forced to use a more detailed representation of the vertical distributions of the gases and scattering properties, then the degeneracies would become more substantial.

This study has shown the benefit of considering a wide spectral range for breaking the degeneracy between different atmospheric parameters. By constraining the overall shape of the infrared spectrum, the $HST$ and $Spitzer$ secondary eclipse data allow us to evaluate the vertical temperature structure and to place crude estimates on the molecular abundances in this hot Jupiter. Studies that focus on smaller wavelength ranges or fewer data points are subject to broader uncertainties than are currently being presented in the literature. Future work will permit an extension of this technique into cross-comparison with transmission spectroscopy, including the visible range (alkali metals, metal oxides, and clouds and hazes) especially at the terminator regions of this planet.

\section*{acknowledgments}

We are grateful to James Cho, Giovanna Tinetti, Suzanne Aigrain, and Sug-Whan Kim for helpful and productive discussions. LNF is supported by a Glasstone Fellowship at University of Oxford. We also acknowledge the support of the UK Science and Technology Facilities Council (STFC). We are extremely grateful to the observational teams responsible for the acquisition and reduction of the datasets described here, without which this analysis would not have been possible.


\label{lastpage}

\end{document}